\documentclass[aps, prl, twocolumn]{revtex4-1}
\usepackage{graphicx}
\usepackage{subfigure}
\usepackage{amsmath,amssymb}
\usepackage{braket}
\usepackage{bm}
\usepackage{color}
\usepackage{latexsym}
\usepackage{amsfonts}
\usepackage[utf8]{inputenc}
\usepackage[colorlinks=true,citecolor=blue,linkcolor=blue]{hyperref}
\usepackage{ulem}

\begin{document}

\title{Emergent devil's staircase without particle-hole symmetry in Rydberg quantum gases with competing attractive and repulsive interactions}

\author{Zhihao Lan, Ji\v{r}\'{i} Min\'{a}\v{r}, Emanuele Levi, Weibin Li and Igor Lesanovsky}
\affiliation{School of Physics and Astronomy, University of Nottingham, Nottingham, NG7 2RD, UK}

\begin{abstract}
The devil's staircase is a fractal structure that characterizes the ground state of one-dimensional classical lattice gases with long-range repulsive convex interactions. Its plateaus mark regions of stability for specific filling fractions which are controlled by a chemical potential. Typically such staircase has an explicit particle-hole symmetry, i.e., the staircase at more than half-filling can be trivially extracted from the one at less than half filling by exchanging the roles of holes and particles. Here we introduce a quantum spin chain with competing short-range attractive and long-range repulsive interactions, i.e. a non-convex potential. In the classical limit the ground state features generalized Wigner crystals that --- depending on the filling fraction --- are either composed of dimer particles or dimer holes which results in an emergent complete devil's staircase without explicit particle-hole symmetry of the underlying microscopic model. In our system the particle-hole symmetry is lifted due to the fact that the staircase is controlled through a two-body interaction rather than a one-body chemical potential.  The introduction of quantum fluctuations through a transverse field melts the staircase and ultimately makes the system enter a paramagnetic phase. For intermediate transverse field strengths, however, we identify a region, where the density-density correlations suggest the emergence of quasi long-range order. We discuss how this physics can be explored with Rydberg-dressed atoms held in a lattice.

 \end{abstract}

\date{\today}

\maketitle

{\it Introduction.}---Systems with long-range interactions can feature intriguing physics that is not necessarily present in their short-range counterparts. For example, classical particles in a one-dimensional (1D) lattice interacting via repulsive infinite-range convex potentials lower their interaction energy by assuming a distribution in space as uniform as possible. Remarkably this property makes the ground state configuration independent of the actual details of the interactions, e.g. the specific power-law of the interaction potential, and leads to the formation
of a so-called generalized Wigner crystal \cite{hubbard, uimin}. Furthermore, the permitted filling fractions of the ground state configuration versus the chemical potential form a fractal curve known as the complete devil's staircase \cite{complete_staircase}.

Recent years have seen a great success in emulating many-body systems with long-range interactions using ultracold gases. Here, crystalline structures which were originally studied in the context of solid state physics have shown to be present also in ensembles of cold trapped ions \cite{ion_staircase}, polar molecules \cite{molecule_staircase1, molecule_staircase2, molecule_staircase3} and gases of Rydberg atoms  \cite{2_melting, lauer2012transport, vermersch_2015, Dislocation_melting}. In particular Rydberg gases have witnessed a recent experimental breakthrough in which the adiabatic preparation of a crystalline state and the onset of a staircase structure were shown \cite{schaus_crystallization_2015}. A case currently much less studied is when long-range interactions feature competing attraction and repulsion. Systems with such interactions are known to exhibit intricate behavior and in fact appear in a number of contexts, e.g., Langmuir monolayers \cite{Langmuir}, classical fluids \cite{Malescio,Archer}, mixtures of colloids and polymers \cite{Sedgwick,Campbell} and ferrofluids \cite{ferrofluids}.

\begin{figure}[!hbp]
\centering
\includegraphics[width=0.88\columnwidth]{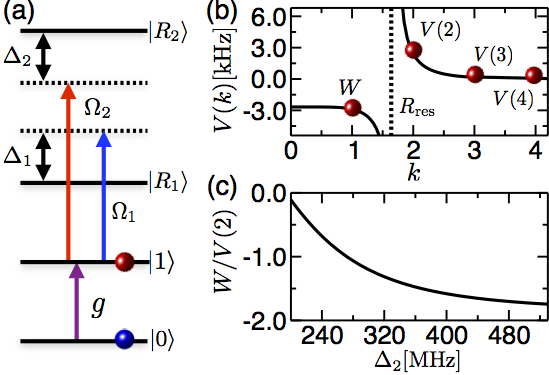}
\caption{(color online) (a) Level scheme. Two ground states $|0\rangle$ and $|1\rangle$ are coupled coherently through a Raman process. The $|1\rangle$ state is weakly dressed to two Rydberg states $|R_1\rangle$ and $|R_2\rangle$ by a blue- and red-detuned lasers. (b) Interaction potential between two atoms in the Rydberg dressed $|1\rangle$ state. Varying the laser parameters and the lattice spacing $a$, we achieve a competing interaction where the NN interaction is attractive and the longer range interactions are repulsive.  The blue-detuned laser induces the two-atom resonant excitation at a distance $R_{\text{res}}$ (dashed line in the figure).  (c) Tuning the competing  interaction by changing the laser detuning $\Delta_2$. In the example, two Rydberg states $|R_1\rangle=|60S\rangle$ and $|R_2\rangle=|70S\rangle$ of Rubidium atoms are considered.  The dispersion constants of the respective Rydberg states are $C_6^{(1)}=140.4\,\text{GHz}\,\mu\text{m}^6$ and $C_6^{(2)}=882.3\,\text{GHz}\,\mu\text{m}^6$. Laser parameters are given in the following. In (b), we assume $\Delta_2=250$ MHz. Other parameters used in plotting (b) and (c) are $\Omega_1=30$ MHz, $\Delta_1=-300$ MHz, $\Omega_2=22$ MHz and $a=2\,\mu$m (see text for details).}
\label{fig:potential}
\end{figure}

In this work we explore such scenario in a cold atomic lattice gas in which interatomic interactions are controlled via a so-called double Rydberg dressing scheme \cite{dressing1,dressing2,dressing3}. This permits the realization of competing nearest-neighbor (NN) attractive and long-range repulsive interactions which can stabilize generalized Wigner crystals and lead to a new mechanism for the formation of a complete devil's staircase. Here the filling fraction is controlled via the NN interaction strength and the emergent devil's staircase is a union of two sub-staircases --- a dimer particle staircase and a dimer hole staircase --- despite of the fact that the microscopic description of the system has no explicit particle-hole symmetry. Going beyond the classical limit we eventually perform a qualitative study of the melting of the crystalline states through quantum fluctuations introduced by a transverse field. In our analysis of the density-density correlations we find signatures of a regime featuring quasi long-range order before a paramagnetic phase is reached.

{\it The system.}--- As shown in Fig.~\ref{fig:potential}(a) we consider atoms in two electronic ground states $|0\rangle$ and $|1\rangle$ (typically Zeeman or hyperfine states) held in a 1D optical lattice. They form a  pseudo-spin $1/2$ particle whose states are (Raman) coupled with strength $g$. One blue-detuned and one red-detuned laser are applied simultaneously to weakly dress the $|1\rangle$ state with two Rydberg $S$-states $|R_1\rangle$ and $|R_2\rangle$, which induces long-range interactions between atoms in the Rydberg dressed $|1\rangle$ state~\cite{henkel10,honer10,cinti10,weibin12}. The double Rydberg dressing in this setting is motivated by recent experimental achievements on Rydberg excitations in lattices or microtraps~\cite{viteau_lattice_2011,anderson_lattice_2012,schaus_observation_2012,li_entanglement_2013,beguin_direct_2013,schaus_crystallization_2015} and laser dressing of ground state atoms to a single Rydberg state~\cite{jau_entangling_2015}.

In the following, we will elaborate on how to build the competing attractive and repulsive interactions in this double dressing scheme. The van der Waals (vdW) interaction of the Rydberg state $|R_j\rangle$ is $C_6^{(j)}/r^6$ with $C_6^{(j)}>0$ the corresponding dispersion constant ($j=1,2$), while the inter-state Rydberg interaction can be neglected when the two states are far separated energetically~\cite{olmos11}. The laser (Rabi frequency $\Omega_j$ and detuning $\Delta_j$) induced interaction is given by $U_j(r)=\tilde{C}_6^{(j)}/(r^6\pm r_j^6)$ with $\tilde{C}_6^{(j)}=r_j^6\Omega_j^4/8|\Delta_j|^3$ and $r_j=(C_6^{(j)}/2|\Delta_j|)^{1/6}$. Here the $+$ and $-$ sign correspond to the red- and blue-detunted laser, respectively. For the dressing by the blue-detuned laser,  $R_{\text{res}}=r_1$ determines the distance of the two-atom resonant excitation when $2\Delta_1+U_1(R_{\text{res}})=0$~\cite{ates12,weibin13} [see Fig. \ref{fig:potential}(b)]. Therefore the resulting interaction $U_1(r)$ is attractive for $r<R_{\text{res}}$ and repulsive when $r>R_{\text{res}}$. The dressing of the red-detuned laser generates a barely repulsive soft-core interaction $U_2(r)$. The overall interaction potential is given by  $V(r)=U_1(r) + U_2(r)$. As depicted in Fig. \ref{fig:potential}(b), it is divided into an attractive (negative) branch and a long-range repulsive  vdW part. Moreover the relative value $W$ at the attractive branch can be tuned by controlling laser parameters, which is illustrated in Fig.~\ref{fig:potential}(c).

In a 1D lattice, the number of lattice sites (red circles) on the attractive branch depends on the length $R_{\text{res}}$. In this work we focus on the case where only the NN interaction is attractive. We parameterize the NN interaction through $W$ and the long-range tail approximately through $V(k)=C_6/(ak)^6$ for each $k \ge2$. Here $C_6\approx C_6^{(1)}+C_6^{(2)}$ and $a$ is the lattice spacing.
The above system is described by the Hamiltonian
\begin{equation}
\hat{H}=g\sum_i \hat{s}^x_i+W\sum_i \hat{n}_i \hat{n}_{i+1}+\sum_{|j-i|>1}\!\!\!V(|i-j|)\,\hat{n}_i\hat{n}_j,
\label{Hamiltonian}
\end{equation}
where $\hat{n}_i=|1 \rangle_i \langle 1|$ and $\hat{s}^x_i=(|1 \rangle_i \langle 0| +|0\rangle_i \langle 1 |)$. Note, that we set $a$ as the length unit and $V(2)$ as the energy unit without loss of generality.

{\it Emergent devil's staircase.}---Let us now determine the ground state configuration of the model Hamiltonian (\ref{Hamiltonian}) in the classical limit ($g=0$). One notable feature is that the crystal formed in the ground state depends on the sign of the NN interaction $W$. In the case of positive $W$ the introduction of an excitation is energetically unfavorable and the ground state of Eq. (\ref{Hamiltonian}) is the empty state ($\prod_i \otimes |0\rangle_i$). On the other hand in the limit of infinite negative $W$ the ground state is the ferromagnetic state ($\prod_i \otimes |1\rangle_i$).
We are interested in how the ground state interpolates between these two cases.
\begin{figure}[!hbp]
\centering
\includegraphics[width=1\columnwidth]{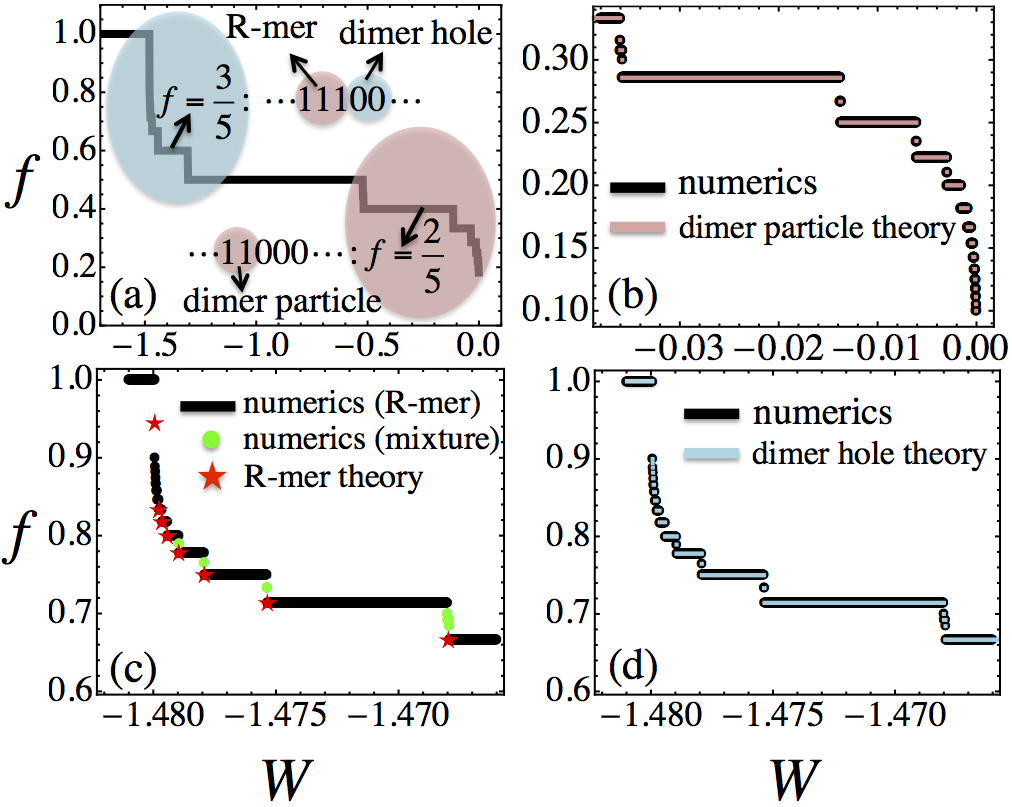}
\caption{(color online) Emergent devil's staircase. (a) Staircase pattern of the filling fraction $f$ versus NN interaction strength $W$. The particle part ($f<1/2$) and the hole part of the staircase ($f>1/2$) are indicated by the brown and blue shaded areas and merge at filling fraction $f=1/2$. (b) Magnified view of the dimer particle staircase. The brown curve is obtained with the effective dimer particle theory of Eq. (\ref{eq:dimerparticle}). The black curve (in all panels) is the numerically calculated staircase taking into account any rational filling fractions $f=p/q$ with $q \leq 20$ where the range of the interaction has been cut off at $k\leq 1000$ lattice sites. (c) At large filling fraction ($f>1/2$) the main plateaus (black) correspond to arrangements of a single kind of $R$-mers. The small intermediate plateaus (green) contain mixtures of $R$-mers and $R+1$-mers.  The red stars correspond to the predicted transition points given by Eq. (\ref{eq:trans cond}). (d) A unified picture for the staircase structure at large fillings is obtained by employing an effective dimer hole theory of Eq. (\ref{eq:dh_theory}) which agrees excellently with the numerical data.}
\label{f2}
\end{figure}

We first perform a numerical analysis of the infinite system. To this end we consider configurations which are periodic with period $q$, i.e. considering rational filling fractions $f=p/q$, with $p \leq q$, and we set the maximal $q=20$. For each value of the interaction $W$ we identify the ground state numerically as the configuration with minimum energy density, i.e., energy of a single period divided by its length. Using this method, we obtain a staircase pattern connecting the empty state ($f=0) $ at $W=0$ with the completely filled state ($f=1$) at $W\simeq-1.48$, depicted in Fig.~\ref{f2}(a). The data show that we never observe isolated excitations, i.e. an atom in state $|1\rangle$ surrounded by atoms in state $|0\rangle$, but rather polymers of neighboring excitations. This is because aggregating at least two excitations is energetically favorable for negative $W$. We find that the excitations arrange in dimers for filling fractions $f<1/2$, which is in agreement with the results of Ref. \cite{dimerstaircase}. Once the threshold of $f=1/2$ is reached the ground state configuration becomes $\cdots 11001100\cdots$. In the region of $f>1/2$ larger polymers form and the numerical results suggest that the ground state exhibits a rather intricate crystalline structure. In the following, we will aim at understanding the staircase analytically in both the regimes of $f<1/2$ and $f>1/2$.

In the regime of  $f<1/2$ --- inspired by the numerical observations that two excitations are always on neighbouring sites --- we formulate an effective dimer particle theory to describe the resulting crystalline structure. We consider the presence of a dimer particle (DP) on sites $(i,i+1)$ as an effective single particle with associated dimer particle number operator $\hat{p}_{i}=\hat{n}_i \hat{n}_{i+1}$. The effective Hamiltonian of the dimers then reads
\begin{gather}
H^{\rm{DP}}=\sum_{i}\sum_{r>2} \mathcal{V}_{\rm p}(r)\hat{p}_{i} \hat{p}_{i+r}+W \sum_i \hat{p}_{i},
\label{eq:dimerparticle}
\end{gather}
where $W$ assumes the role of a chemical potential for the dimers.
Here $\mathcal{V}_{\rm p}(r)=2V(r)+V(r+1)+V(r-1)$ is the interaction potential between two dimers separated by $r$ sites, which is convex.
Note, that for $f<1/2$ the convexity of the potential ensures that no neighbouring dimers are allowed. We can now employ a method used by Hubbard \cite{complete_staircase} to identify stability regions of different densities of dimers with respect to $W$. The result is plotted in Fig. \ref{f2}(b). One observes a remarkable agreement between the dimer model and the numerical result by solving the full Hamiltonian.

In the regime of  $f>1/2$ --- unlike the traditional staircase where the structure of the staircase in this hole sector is trivial, a detailed study must be carried out since there is no explicit particle-hole symmetry in our system (\ref{Hamiltonian}). From the numerical data we find that in this regime excitations form polymers constituted of $R$ neighboring excitations (so-called $R$-mers). We find regions in which only one kind of $R$-mer is stable, and regions in which different kinds of $R$-mers coexist. The former correspond to the wider plateaus in the staircase [black data points in Fig. \ref{f2}(c)]. The latter --- ``mixed'' cases ---  correspond to the narrow regions interpolating between two wider plateaus [green data points in Fig. \ref{f2}(c)].

As a first approximation, we can construct  an $R$-mer theory to describe the staircase as the dominant plateaus correspond to different kinds of $R$-mers (without mixtures). Our aim is to determine when the crystal changes from $R$-mer to $R+1$-mer as we change $W$, which is achieved by examining the stability of $R$-mers for a given $W$. Considering an $R$-mer as a single particle, we define the effective filling fraction of $R$-mers as $f_{\rm eff}=P/Q$, which is related to the real filling fraction $f=p/q$ through $f_{\rm eff}=p/[R(q-p)+p]$. In the following we focus on filling fractions with $f_{\rm eff}=1/Q$. The energy of an $R$-mer in the thermodynamic limit reads
\begin{equation}
 E^{\mathrm{tot}}_R=E^{\mathrm{b}}_R+\sum_{l=1}^{\infty}E_R^{\mathrm{int}}(lu)
 \end{equation}
where $E^{\mathrm{b}}_R= (R-1)W+\sum_{l=1}^{R-1}(R-l)V(l)$ is the bond energy of the $R$-mer, which is associated to the mutual interactions of its internal excitations. $E_R^{\mathrm{int}}(x)=R V(x) + \sum_{l=1}^{R-1}(R-l)\left[V(x+l)+V(x-l)\right]$ is the interaction energy between two $R$-mers separated by $x$ ($x=lu$), where $u$ is the length of the unit cell.

When the crystal changes from $R$-mer to $R+1$-mer, the corresponding unit cell length changes from $u=Q+R-1$ (for $R$-mer) to $u'=Q'+R$ (for $R+1$-mer).
The infinite chain can be divided in equal periods on which either $u'$ $R$-mers or $u$ $R+1$-mers  are disposed, so that we can compute the transition point by solving
\begin{equation}
  u E^{\mathrm{tot}}_{R+1}=u'E^{\mathrm{tot}}_{R}.
  \label{eq:trans cond}
  \end{equation}
We observe numerically that in the regions in which only one kind of $R$-mer is stable the $R$-mers are homogeneously distributed and separated by two holes, giving $f_{\rm{eff}}=1/3$ and therefore $Q=Q'=3$.
This numerical observation is confirmed analytically in \cite{suppl}, where we show that any other filling fraction makes the resulting $R+1$-mer arrangement unstable.
The comparison of $W$ obtained from the simple theoretical model Eq. (\ref{eq:trans cond}) with the numerical data is shown in Fig. \ref{f2}(c). One can see that the transition values of $W$ are in excellent agreement.

The presence of two neighboring holes in each unit cell stimulates us to search for an effective theory of dimers of holes (DH). In analogy to the dimers of particles Eq. (\ref{eq:dimerparticle}), the effective Hamiltonian reads
\begin{gather}
H^{\mathrm{DH}}=\sum_{i}\sum_{r>2}\mathcal{V}_{\rm h}(r)\hat{h}_{i} \hat{h}_{i+r}-\mu_\mathrm{h}\sum_i \hat{h}_{i} , \label{eq:dh_theory}
\end{gather}
where $\mathcal{V}_{\rm h}(r)=\mathcal{V}_{\rm p}(r)$ is the interaction energy between dimer holes,  $\hat{h}_i = (1-\hat{n}_i)(1-\hat{n}_{i+1})$ is the dimer hole number operator and $\mu_\mathrm{h}=3W+4\sum_{j=2}^{\infty}V(j)$ can be regarded as a chemical potential. We can again study the stability regions with a method used by Hubbard \cite{complete_staircase} as the interaction is convex. In Fig. \ref{f2}(d) we report the filling fraction of excitations found in this way, and we compare them with the numerical results of the full Hamiltonian, finding a perfect agreement in the region $f>1/2$.
Note by setting the chemical potential $\mu_h=0$, no holes are present, and a single polymer extends over the whole system corresponding to the aforementioned ferromagnetic state.
We can then identify the transition value of $W$ to the ferromagnetic state exactly, i.e., $W_c=-4/3 \sum_{j=2}^{\infty}V(j)=-1.47994$. Note that the emergent staircase is the union of two sub-staircases and is complete due to the fact that the long-range interactions $\mathcal{V}_{\rm h}(r)$ and $\mathcal{V}_{\rm p}(r)$ in the respective effective models are convex~\cite{complete_staircase}.

\begin{figure}[!hbp]
\centering
\includegraphics[width=1\columnwidth]{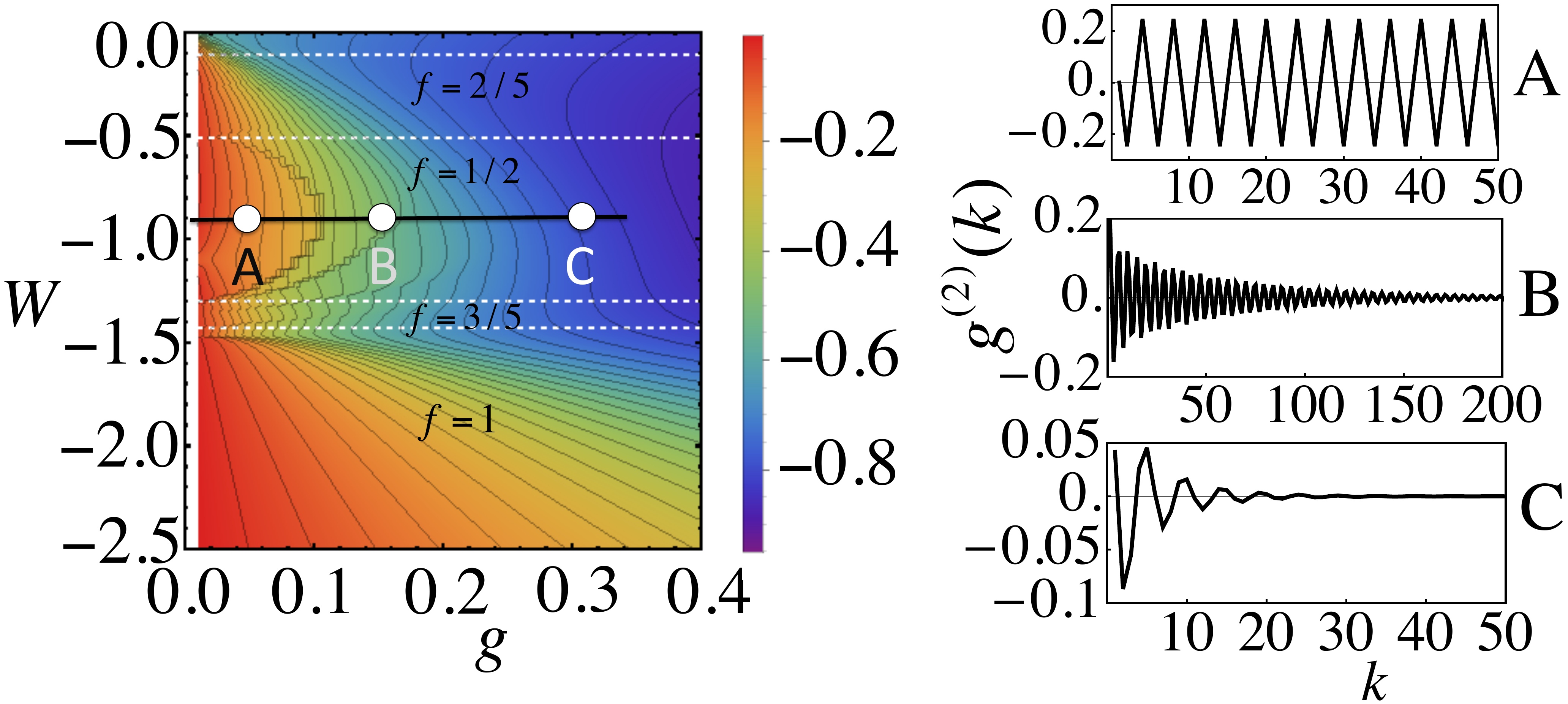}
\caption{(color online) Quantum melting of the dimer crystal with increasing transverse field strength $g$. Left panel shows the transverse magnetization $\sum_i \langle \hat{s}_i^x\rangle /L$ as function of $W$ and $g$, where the simulation was done for $L=60$ lattice sites with open boundary conditions and the range of the interactions has been truncated to 10 lattice sites. The four largest stability regions $f=1, f=3/5,f=1/2,f=2/5$ are marked by dashed lines, see also Fig. \ref{f2}. Right panel shows the  iDMRG calculations of the density-density correlation function $g^{(2)}(k)$ vs. $k (\ge1)$ for three different values of $g=(0.05,0.15,0.3)$  at $W=-0.91$ corresponding to the points A,B,C in the left panel (see text for details).}
\label{f3}
\end{figure}

{\it Melting of the staircase.}--- In the case of a long-range repulsive interaction, it was shown in \cite{2_melting} that  the introduction of a quantum term leads to a two-stage melting process when the excitations are dilute. For weak transverse fields the crystalline structure melts into a floating solid which will eventually melt into a paramagnet for higher values of  $g$. Here we explore the qualitative features of the melting process in the context of competing attractive and repulsive interactions for a relatively dense crystal. We perform numerical studies by means of Matrix Product States implementations of the Density Matrix Renormalization Group (DMRG) \cite{DMRG} and its infinite version iDMRG \cite{iDMRG}. The results are presented in Fig. \ref{f3}. The left panel shows the transverse magnetization $\sum_i \langle \hat{s}_i^x\rangle /L$ as function of $W$ and $g$, where $L$ is the length of the chain.  The transverse magnetization increases with increasing $g$. Typically crystals in the middle of a plateau are more robust against the quantum fluctuation, which leads to a series of lobes similar to the one found in \cite{2_melting,igor}.

Here we study the melting of the $f=1/2$ lobe which has large stability region and is thus naturally easier to access experimentally (see \cite{suppl} for a detailed discussion of the melting process for lobes with different filling fraction). We use the density-density correlation function $g^{(2)} (k)=\langle \hat{n}_i \hat{n}_{i+k} \rangle-\langle \hat{n} \rangle ^2$ (which is independent of the initial site $i$ for an infinite system) to characterize the melting process along the black line in Fig. \ref{f3}. Performing a numerical simulation for vanishingly small values of $g$ (close to the classical limit) is challenging due to the presence of quasi translational invariance which leads to a highly degenerate ground state.
We consider $g=0.05$ (see point A in Fig. \ref{f3}), which allows the degeneracy to be slightly lifted while still inside the lobe. We find a periodic correlation function $g^{(2)}(k)$ characteristic for a crystalline state. In our case ($f=1/2$) the ground state is quasi-four-fold degenerate and for small values of $g$ has the structure $| \psi_g \rangle \propto (|..1100..\rangle+|..0110..\rangle+|..0011..\rangle+|..1001..\rangle)/2+O(g^2)$ that allows us to derive the approximate expression $g^{(2)}(k)=\cos\left(k\pi/2\right)/4+O(g^2)$ which reproduces remarkably well the results shown for point A in Fig. \ref{f3}.

When $g$ is sufficiently large, the ground state of the system is in a paramagnetic phase and the envelope of $g^{(2)}(k)$ shows an exponential decay (point C in Fig. \ref{f3}). By considering intermediate values of $g$ we study the melting of the crystalline phase and find that in this regime  the correlations are significantly longer ranged. In the right panel of Fig. \ref{f3}, we show a typical result with bond dimensions of 100 at $g=0.15$ (point B in Fig. \ref{f3}). The  correlations in the vicinity of the point B show similar behavior suggesting that there is a region with quasi long-range order extending from $g\simeq 0.1$  to $ g\simeq 0.2$.

{\it Summary and Outlook.}---In conclusion, we have shown that a long-range interaction with competing attractive and repulsive parts can result in  a new mechanism to form a complete devil's staircase. In particular we have considered the case of attractive NN interactions which lead to a dimer staircase without manifest particle-hole symmetry. This suggests an interesting way to control the structure of the hole sector of the staircase by tailoring the repulsive tail. For example, one will obtain a trimer-hole staircase through replacing the longer ranged interaction part from the vdW shape ($1/r^6$) to a dipolar interaction potential ($1/r^3$). Moreover, since in the outlined double Rydberg dressing scheme the range of the attractive branch of the potential can be tuned freely, and therefore be extended over more lattice spacings, we expect that more exotic polymer staircases can be explored.
 \begin{acknowledgments}
\textit{Acknowledgements.---} The research leading to these results has received funding from the
European Research Council under the European Union's Seventh Framework
Programme (FP/2007-2013) / ERC Grant Agreement No. 335266 (ESCQUMA), the
EU-FET Grant No. 512862 (HAIRS), the H2020-FETPROACT-2014 Grant No.
640378 (RYSQ), and EPSRC Grant No. EP/M014266/1. WL is supported through
the Nottingham Research Fellowship by the University of Nottingham.

\end{acknowledgments}

\newpage
\begin{widetext}
\section{Supplementary Material}

\section{I: lifetime of the dressed ground state atoms}
Here we briefly discuss the lifetime of the Rydberg dressed ground state atoms~\cite{henkel10}.  Lifetimes in the Rydberg states $|60S\rangle$ and $|70S\rangle$ are 252 $\mu$s and 410 $\mu$s~\cite{beterov_2009}, correspondingly. The lifetime in the dressed state $|1\rangle$  is amplified by a factor of $(2\Delta_{1,2}/\Omega_{1,2})^2 \simeq 400$ due to the weak mixing with the Rydberg states. With the parameters in the caption of Fig. 1 in the main text, we find the effective lifetime is about $0.068$ second, which is far longer than other time scales in the Hamiltonian.

\section{II: Analytical calculation of the transition point between two R-mers}
In Eq.(4) in the main text we studied the transition between a gas of $R$-mers and $R+1$-mers assuming the effective filling fraction $f_{\mathrm{eff}}=1/3$ for both gases.
Here we want to justify analytically that assumption.
As it stands Eq.(4) does not constrain the effective filling fractions of the two gases, which can in principle be different.
The crucial constraint comes from the requirement that at the transition point both the filling fractions of the gas of $R$- and $R+1$-mers become unstable.
For the gas of $R$-mers, the boundaries of the stability region of a certain filling fraction can be determined from the energetics: we call $W^{\pm}$ the points at which a gas with one more/less $R$-mer becomes energetically favorable. 
An analytic expression for $f_{\mathrm{eff}}=1/Q$ was found in \cite{complete_staircase,Levi_2015_1,Levi_2015_2} and it reads
\begin{eqnarray}
 W^{-} &=& -\frac{1}{R-1} \left[ E^{\rm b,int}_R + \sum_{l=1}^\infty r_l E^{\rm int}_R(r_l-1) - (r_l-1)E^{\rm int}_R(r_l) \right] \label{eq:V1-} \\
 W^{+} &=& -\frac{1}{R-1} \left[ E^{\rm b,int}_R + \sum_{l=1}^\infty (r_l+1) E^{\rm int}_R(r_l) - r_l E^{\rm int}_R(r_l+1) \right], \label{eq:V1+}
\end{eqnarray}
where $E^{\rm b,int}_R = \sum_{l=1}^{R-1} (R-l)V_l$ is the repulsive interaction energy between the constituents of a single $R$-mer and $r_l=l(R+Q-1)$ is the distance between first spins of $l$-th nearest $R$-mers.
The expressions in Eqs.(\ref{eq:V1-},\ref{eq:V1+}) are derived under the assumption that the interaction between the $R$-mers is repulsive, i.e. in our case it breaks down when the configuration of the gas can allow two $R$-mers sitting next to each other, which happens for high filling fractions. 
We need then to restrict to the case $r_1 > R+1$ as for $r_1 = R+1$ the distances $r_1 - 1 = R$ in Eq.(\ref{eq:V1-}) would correspond to adjacent $R$-mers. We plot the devil's staircase for different $R$-mers in Fig. \ref{fig:Rmers}.
\begin{figure}[!h]
\centering
\includegraphics[width=0.5\columnwidth]{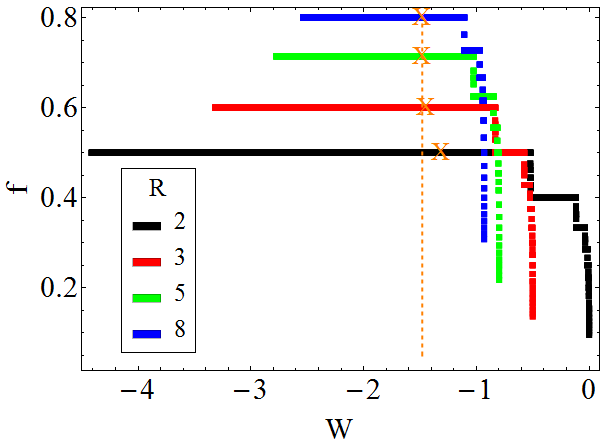}
\caption{(color online) The devil's staircase for a gas of $R$-mers. The black, red, green and blue staircases correspond to the gas of dimers, trimers, pentamers and octamers respectively. The orange crosses indicate the transition points between different polymers (e.g. the transition point between dimers and trimers is represented by the orange cross at filling fraction 0.5). As $R$ is increasing the transition value $W$ quickly approaches the asymptotic value $W(R=\infty) \approx -1.48$, which is indicated by the orange dashed line. The parts of the staircase for higher filling fractions (after the large plateaus as $W$ is decreased) are not shown as they are not captured by our simple theory (see text for details).}
\label{fig:Rmers} 
\end{figure}

With the stability region at hand one can find the transition points between different kinds of $R$-mers as follows: firstly we find the regions in $W$ over which two different $R$-mers are stable, and extract the effective filling fractions of the two gases.
Second we plug the extracted filling fractions in Eq.(4) to find the transition point. 
This point corresponds to a real transition only if it falls in the overlap of the region of stability of the two gases.

To illustrate this with an example, let us consider e.g. a filling fraction $f=0.5$ which admits both dimers (see Fig. \ref{fig:Rmers}, black line, $f_{\rm eff} = 1/3$) and trimers (red line, $f_{\rm eff}=1/4$). The condition Eq.(4) then yields $W = - 1.6858$, which is inconsistent with the given stability region $W \in \left[-0.8319, - 0.5759 \right]$ of trimers. In this way one can show that as $W$ is decreased, the chain is filled with dimers only down to $W = -1.3066$, where the first transition to trimers occurs (orange cross at $f=0.5$ in Fig. \ref{fig:Rmers}). Following this logic it is then straightforward to show, that all the remaining transitions between $R$-mers and $R+1$-mers occur at $f_{\rm eff} = 1/3$ and quickly approach the asymptotic value $W(R \rightarrow \infty) \cong -1.4799$. The values of $W$ for some of the transitions are shown in Table \ref{tab:trans points} and plotted in Fig. 2c in the main manuscript.

\begin{table}[t]
\begin{center}
  \begin{tabular}{ | c | c | }
    \hline
    $R/(R+1)$ & $W$ \\ \hline
    2/3 & -1.3066  \\ 
    3/4 & -1.4414  \\
    4/5 & -1.4680  \\
    5/6 & -1.4754  \\
    6/7 & -1.4799  \\
    7/8 & -1.4789  \\
    8/9 & -1.4794  \\
    ... & ... \\
    $\infty$ & -1.4799 \\
    \hline    
  \end{tabular}
  \caption{Values of $W$ where a transition between $R/(R+1)$-mers occurs, determined from Eq.(4) with $f_{\rm eff}=1/3$.}
  \label{tab:trans points}
\end{center}
\end{table}

\section{III: quantum melting of the $f=2/5$ lobe}

\begin{figure}[!h]
\centering
\includegraphics[width=0.8\columnwidth]{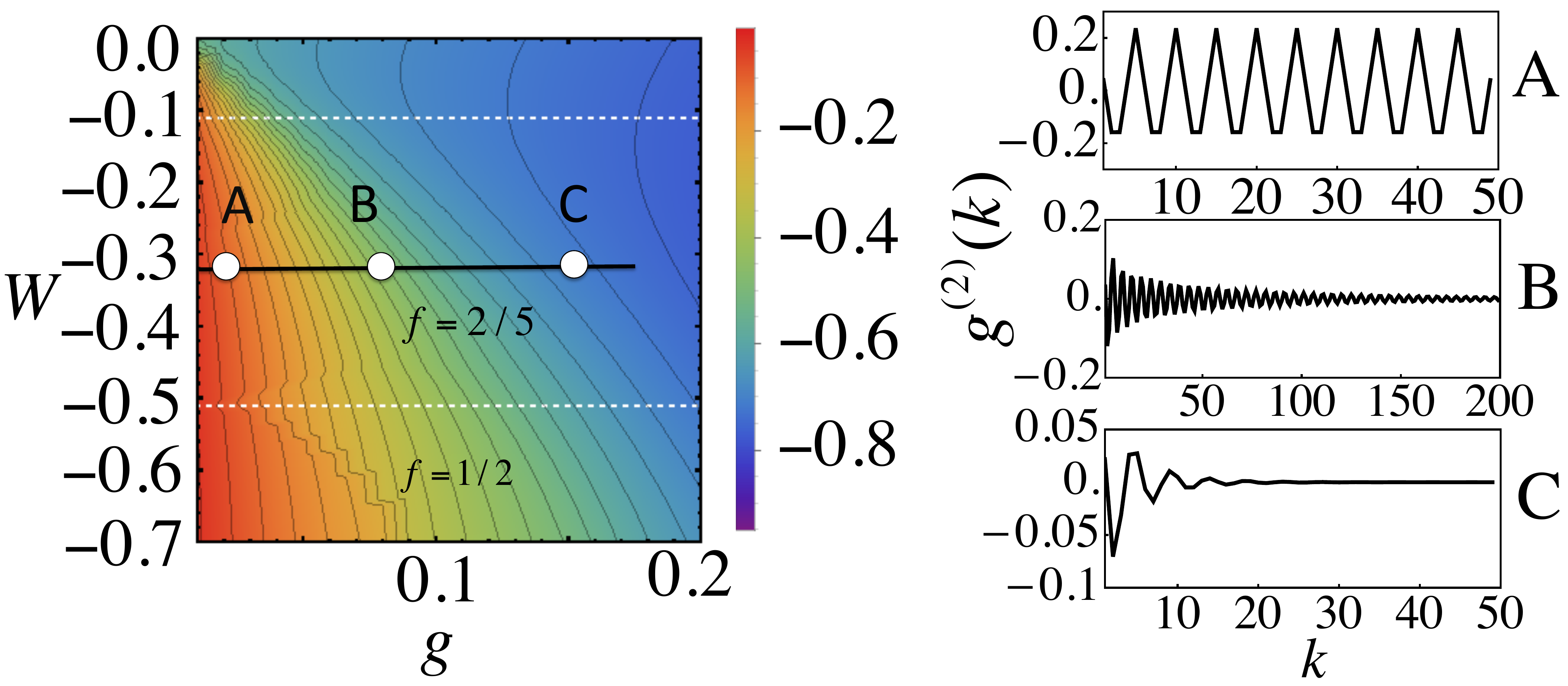}
\caption{(color online) Quantum melting of the $f=2/5$ lobe. The left panel shows the transverse magnetization $\sum_i \braket{\hat{s}^x_i}/L$ in the vicinity of $f=2/5$ (see the caption of Fig. 3 in the main text for further details). The right panel shows the  iDMRG calculations of the density-density correlation function $g^{(2)}(k)$ vs. $k (\ge1)$ for three different values of $g=(0.02,0.08,0.15)$ at $W=-0.32$ corresponding to the points A,B,C in the left panel.}
\label{fig:sf2} 
\end{figure}

In the main text, we have focused on the quantum melting of the $f=1/2$ lobe, which is the largest lobe between $f=0$ and $f=1$. However, the repulsive van der Waals tail produces an intricate staircase-like structure of the crystalline ground states (see Fig.1 of the main text). To understand the effects of the repulsive tail, as a first approximation, we could keep only the nearest-neighbour attraction and the next-nearest-neighbour repulsion, and as such the model reduces to the well-known ANNNI (axial next-nearest-neighbour Ising) model \cite{Ising}, for which only three possible ground states exist at $g=0$, i.e., the $f=0$, $f=1/2$ and $f=1$ phases. The transition from the phase of $f=0$ to  that of $f=1/2$ happens at $W=0$ as explained in the main text, and the transition from the $f=1/2$  phase to that of $f=1$ happens at $W/V(2)=-4/3$ \cite{Ising}. Now if one introduces the seemingly small third nearest-neighbour repulsion $V(3)$, which is equal to $V(2)/64$ due to the $1/r^6$ law, two additional new phases appear at filling fractions $f=2/5$ and $f=3/5$. 
The filling fraction $f=2/5$ is stabilized for  $W \in [-0.518, -0.116]$ which is roughly half of the width of the stability region of $f=1/2$ ($W \in [-1.306, -0.522]$). Importantly, the filling fraction $f=2/5$ is stabilized due to the interaction terms going beyond next-to-nearest neighbour. For interaction with the van der Waals tails, the physics in the sector with filling fraction $f\le 1/2$ is dominated by pure dimers and the effective interaction between these dimers is given by the effective potential $\mathcal{V}_{\rm p}$ rather than by the van der Walls potential itself.

In Fig. \ref{fig:sf2} we show the melting process of the $f=2/5$ lobe. The qualitative features are analogous to those for $f=1/2$ studied in the main text. Inside the $f=2/5$ lobe, the crystalline structure is well preserved as indicated by the periodic behaviour of the $g^{(2)}(k)$ correlation function (point A in Fig.\ref{fig:sf2}). The ground state wave function at small $g$ is given by $| \psi_g \rangle \propto (|..11000..\rangle+|..01100..\rangle+|..00110..\rangle+00011..\rangle+|..10001..\rangle)/\sqrt{5}+O(g^2)$, from which one can calculate $g^{(2)}(k=1,2,3,4,5\cdots)=(1/25, -4/25, -4/25, 1/25, 6/25 \cdots)$ which matches perfectly the numerical results. Experimentally, one can distinguish the different crystalline orders at small $g$ by measuring the different periodic behaviours of $g^{(2)}(k)$. At intermediate $g$, one observes the quasi long-range order indicated by the algebraic decay of the correlations (point B in Fig.\ref{fig:sf2}). Eventually the crystal melts into a paramagnetic phase with exponentially decaying correlations (point C in Fig.\ref{fig:sf2}).

\end{widetext}

\end{document}